\documentclass[aps,reprint,amsmath,amssymb]{revtex4-2}
\usepackage[dvipsnames]{xcolor}
\usepackage[utf8]{inputenc}
\usepackage[T1]{fontenc}
\usepackage{bm,physics,siunitx}
\usepackage{graphicx}
\usepackage[colorlinks]{hyperref}
\usepackage{cleveref}
\usepackage{microtype}
\usepackage{enumitem}
\everymath{\displaystyle}
\hypersetup{linkcolor=ForestGreen,
            citecolor=BrickRed,
            urlcolor=RoyalBlue}
\newcommand\scalemath[2]{\scalebox{#1}{\mbox{\ensuremath{\displaystyle #2}}}}
\setlist{noitemsep}
\setlist{nosep}

\begin{document}

\title{A photon density wavefunction}

\author{St\'ephane Virally}
\email{stephane.virally@polymtl.ca}
\affiliation{Polytechnique Montr\'{e}al, Engineering Physics, 2500 chemin de Polytechnique, Montr\'eal, QC H3Y2T9, Canada}

\date{\today}

\begin{abstract}
Maxwell's equations in the vacuum can be formally cast in the form of a Schr\"odinger equation. The vector to which this equation directly applies is not a regular wavefunction: its amplitude squared is not a probability density but the expected energy density of the field. Since clicks on a photodetector can be observed, there should exist a different wavefunction, still derived from the EM field, whose amplitude squared is an expected photon density. Mandel proposed the second quantized version of the amplitude of such a wavefunction, but did not link it directly to the EM field. We show how this can be accomplished.
\end{abstract}

\maketitle
{\bf Introduction.}
Quantum optics, and particularly the second quantization step~\cite{walls2008quantum}, is often performed in the frequency domain, where the notion of energy and the Hamiltonian play a central role. In the time domain, quasi-instantaneous observables such as clicks on a photodetector are a more natural fit. The notion of a wavefunction whose amplitude squared would predict the rate of photodetections in a small volume has been the topic of many discussions but is not yet settled. Glauber pointed to the fact that a click on a photodetector is best understood in a wideband setting~\cite{glauber2006hundred}[p.~1271], and Mandel explored the properties of a ``detection operator''~\cite{mandel1966configuration}, a broadband annihilation operator $\hat{\phi}(\bm{r},t)$ better localized in real spacetime than in energy/momentum space. The photon number observable that follows, $\hat{\phi}^\dagger\,\hat{\phi}$, when integrated over a volume $V$, corresponds to the number of clicks on a photodetector inside that volume, as long as the largest wavelength in the detector bandwidth is smaller than the smallest dimension of the probed volume. Mandel did not, however, link his detection operator directly to any observable of the electromagnetic field.

The main difficulty stems from the fact that the electromagnetic (EM) field is essentially a carrier of energy, not particles. So far, all attempts at finding a wavefunction for photons starting from Maxwell's equations have lead to states and operators corresponding to a density of energy rather than a density of particles. Prominently Bia\l{}ynicki-Birula~\cite{BialynickiBirula1996_OnTheWaveFunctionOfThePhoton} introduced a Schr\"odinger-like equation that applies to the Riemann-Silberstein vector $\textstyle{\bm{\mathcal{D}}/\sqrt{2\epsilon_0}+i\,\bm{\mathcal{B}}/\sqrt{2\mu_0}}$~\cite{silberstein1914theory,bialynicki1996photon}, whose squared amplitude is the energy density of the electromagnetic (EM) field, not a density of particles. The second-quantized version of the amplitude of this wavefunction, $\hat{\psi}(\bm{r},t)$, is distinct from Mandel operator: $\hat{\psi}^\dagger\,\hat{\psi}$ is the Hamiltonian, not the number operator. The article in which this second-quantized form was introduced in fact argues that for this reason, energy is the natural way to localize the field~\cite{sipe1995photon}.

This work showcases a direct link between Maxwell's equations and a photon density wavefunction, through a form of Schr\"odinger's equation well suited for massless spin-1 particles such as photons, and shown to be equivalent to Maxwell's equations. It also demonstrates that energy and particles are in fact exactly on the same footing when it comes to localization. It does so in the following manner: first-order Maxwell's equations are formally cast into Schr\"odinger form, with one solution the Riemann-Silberstein vector; a causal transform~\cite{Virally2019_UnidimensionalTimeDomainQuantumOptics} of this vector converts it into another solution, whose amplitude squared is this time the expected photon density. The second-quantized version of this wavefunction's amplitude is shown to be Mandel's operator.

\vspace{5pt}{\bf Bia\l{}ynicki-Birula's equation.}
In free space, Maxwell's equations describing the unitary evolution in real space and time of the electromagnetic (EM) field can be written using the transverse fields $\bm{\mathcal{D}}$ (electric displacement field) and $\bm{\mathcal{B}}$ (magnetic field) as~\cite{griffithsEM}
\begin{center}
\vspace{-10pt}
\begin{subequations}
    \begin{minipage}{0.24\textwidth}
        \begin{equation} \label{McrossD}
        \curl\bm{\mathcal{D}}=-\epsilon_0\pdv{\bm{\mathcal{B}}}{t},
    \end{equation}
\end{minipage}%
\begin{minipage}{0.24\textwidth}
    \begin{equation} \label{McrossB}
        \curl\bm{\mathcal{B}}=\mu_0\pdv{\bm{\mathcal{D}}}{t},
    \end{equation}
\end{minipage}%
\end{subequations}
\end{center}
where $\epsilon_0$ is the permittivity and $\mu_0$ the permeability of free space.

Interestingly, we can rewrite $\curl\equiv-i\,\bm{L}\cdot\bm\nabla$, where $\bm{L}=\qty(L_x,L_y,L_z)$ is the hermitian form of the matrix representation of $\mathfrak{so}(3)$, the Lie algebra associated with spin-1 particles (see Appendix~\ref{ObviousL}). Using the definition of the canonical momentum operator $\hat{\bm{p}}=-i\hbar\bm{\nabla}$~\cite{sakuraiQM}, \cref{McrossD,McrossB} can be formally cast as Schr\"odinger's equation~\cite{schrodinger1926,sakuraiQM}
\begin{equation}
    i\hbar\pdv{t}\ket{\bm\Psi}=\bm{L}\cdot\hat{\bm{p}}\,c\ket{\bm\Psi},\label{SPsi}
\end{equation}
with $c=1/\sqrt{\epsilon_0\mu_0}$ the speed of light in the vacuum and
\begin{equation}
    \ket{\bm\Psi}=\frac{\bm{\mathcal{D}}}{\sqrt{2\epsilon_0}}+i\,\frac{\bm{\mathcal{B}}}{\sqrt{2\mu_0}},\label{Psi}
\end{equation}
the Riemann-Silberstein vector. This is, in slightly more compact form, the result obtained by Bia\l{}ynicki-Birula in 1996. It features a linear dispersion relation characteristic of massless particles and a tensor form with the right structure for spin-1 particles.

\vspace{5pt}{\bf A photon density wavefunction.}
\Cref{SPsi} is obtained from Maxwell equations in such a way that \cref{Psi} is an immediate solution. Unfortunately, it is not the result we are looking for: the squared amplitude of the Riemann-Silberstein vector is an energy density, not a photon density.

To obtain another solution with the right properties, we start by defining a transverse potential vector in the Coulomb gauge (see Appendix~\ref{ObviousPotential}). The transverse electric displacement and transverse magnetic field are derived by simple differentiation of this vector potential with respect to time and space. Writing the Riemann-Silberstein vector as a sum of a positive-frequency and negative-frequency part, $\bm\Psi=\bm\Psi^{(+)}+\bm\Psi^{(-)}$, we get
\begin{equation}
    \begin{split}
    \braket{\bm{r},t}{\bm\Psi^{(\pm)}}&=\qty(\pm i)\int_{\mathbb{R}^3}\frac{\dd^3\bm{k}}{(2\pi)^{3/2}}\;\sum_{\sigma\in\qty{+,-}}\\&
    \sqrt{\hbar\,\abs{\bm{k}}c}\;A^{(\pm)}_\sigma(\bm{k})\,e^{i(\bm{k}\cdot\bm{r}\mp\abs{\bm{k}}ct)}\;\tilde{\bm{u}}_\sigma(\bm{k}),\label{e0Em0H}
    \end{split}
\end{equation}
where $\sigma$ is the helicity, $\textstyle{A^{(\pm)}_\sigma(\bm{k})}$ are complex unitless amplitudes and $\tilde{\bm{u}}_\sigma(\bm{k})$ complex unit vectors that verify the conditions
\begin{equation}
    \begin{matrix}
    A^{(-)}_\sigma(\bm{k})=\qty[A^{(+)}_\sigma]^*(\bm{k});
    &A^{(+)}_\sigma(\bm{-k})=A^{(-)}_{-\sigma}(\bm{k});\\
    \tilde{\bm{u}}_\sigma(\bm{k})\cdot\tilde{\bm{u}}^*_{\sigma'}(\bm{k})=\delta_{\sigma,\sigma'};&\tilde{\bm{u}}_\sigma(\bm{-k})=\tilde{\bm{u}}_{-\sigma}(\bm{k}).
    \end{matrix}\label{anti}
\end{equation}
Here, we can trace the fact that $\textstyle{\norm{\braket{\bm{r},t}{\bm\Psi}}^2}$ is a density of energy to the $\sqrt{\hbar\,\abs{\bm{k}}c}$ factor in the integral.

Following an idea developed recently~\cite{Virally2019_UnidimensionalTimeDomainQuantumOptics}, we can remove this factor by applying the causal transform (see more in Appendix~\ref{Transforms})
\begin{equation}
    \mathcal{T}_+[\bm{\mathcal{F}}](t)=\sqrt{\frac{2}{\hbar}}\int_{0}^{+\infty}\dd\tau\;\frac{\bm{\mathcal{F}}(t-\tau)}{\sqrt{\tau}}.\label{T}
\end{equation}
Applied to~\cref{e0Em0H}, it removes the $\sqrt{\hbar\,\abs{\bm{k}}c}$ factor in the integral and adds a $\pm\pi/4$ phase to the positive and negative frequency parts,
\begin{equation}
    \begin{split}
    &\mel{\bm{r},t}{\mathcal{T}_+}{\bm\Psi^{(\pm)}}\equiv\braket{\bm{r},t}{\bm\Phi^{(\pm)}}=\\&\qty(\pm i)\int_{\mathbb{R}^3}\frac{\dd^3\bm{k}}{(2\pi)^{3/2}}\;\sum_{\sigma}
    A^{(\pm)}_\sigma(\bm{k})\,e^{i(\bm{k}\cdot\bm{r}\mp\abs{\bm{k}}ct\pm\pi/4)}\;\tilde{\bm{u}}_\sigma(\bm{k}).\label{Phipm}
    \end{split}
\end{equation}
Since $\mathcal{T}_+$ commutes with space and time derivatives, the vector $\ket{\bm\Phi}$ also verifies~\cref{SPsi}. Additionally, its amplitude squared is now a density of particles. It is in fact the first-quantized version of Mandel operator, as we show below.

The wavefunction $\ket{\bm\Phi}$ thus verifies the natural Schr\"odinger equation for massless spin-1 photons, directly derived form Maxwell equations, and its squared amplitude is a density of particles. This is our main result.

\vspace{5pt}{\bf Second quantization.}
The second-quantization step is here straightfoward and does not require any quantization volume. It is enough to swap the unitless amplitudes of~\cref{e0Em0H,Phipm} for ladder operators verifying the usual commutation relations (see Appendix~\ref{ObviousCommutation}):
\begin{equation}
    \begin{matrix}
        A^{(+)}_\sigma(\bm{k})\rightarrow\hat{a}_{\sigma,\bm{k}};
        &A^{(-)}_\sigma(\bm{k})\rightarrow\hat{a}^\dagger_{\sigma,\bm{k}}.
    \end{matrix}\label{Q2}
\end{equation}
The conditions~\eqref{anti} imply that $\hat{a}_{\sigma,-\bm{k}}=\hat{a}^\dagger_{-\sigma,\bm{k}}$, which simply states that photons with opposite $\bm{k}$ vectors and opposite helicities are antiparticles of one another.

The second-quantized versions of $\bm\Psi^{(+)}_\sigma$ and $\bm\Phi^{(+)}_\sigma$ are the annihilation-like,
\vspace{-5pt}
\begin{equation}
    \widehat{\bm\psi}_\sigma(\bm{r},t)=\int_{\mathbb{R}^3}\frac{\dd^3\bm{k}}{(2\pi)^{3/2}}\,\sqrt{\hbar\,\abs{\bm{k}}c}\;
    \hat{a}_{\sigma,\bm{k}}\,e^{i(\bm{k}\cdot\bm{r}-\abs{\bm{k}}ct)}\;\tilde{\bm{u}}_\sigma(\bm{k});\label{Sipe}
\end{equation}
\vspace{-10pt}
\begin{equation}
    \widehat{\bm\phi}_\sigma(\bm{r},t)=\int_{\mathbb{R}^3}\frac{\dd^3\bm{k}}{(2\pi)^{3/2}}\;
    \hat{a}_{\sigma,\bm{k}}\,e^{i(\bm{k}\cdot\bm{r}-\abs{\bm{k}}ct+\pi/4)}\;\tilde{\bm{u}}_\sigma(\bm{k}),\label{Mandel}
\end{equation}
and their negative frequency counterparts are creation-like (see Appendix~\ref{ObviousCreation}).

The second operator is a wideband photon annihilator equal to Mandel detection operator, up to the $\pi/4$ phase.

The Hamiltonian and number operator now can be written
\begin{equation}
    \widehat{H}(t)=\int_{\mathbb{R}^3}\dd^3\bm{r}\;\sum_{\sigma}\;\widehat{\bm\psi}^\dagger_\sigma(\bm{r},t)\cdot\widehat{\bm\psi}_\sigma(\bm{r},t);\label{H}
\end{equation}
\vspace{-10pt}
\begin{equation}
    \widehat{N}(t)=\int_{\mathbb{R}^3}\dd^3\bm{r}\;\sum_{\sigma}\;\widehat{\bm\phi}^\dagger_\sigma(\bm{r},t)\cdot\widehat{\bm\phi}_\sigma(\bm{r},t).\label{N}
\end{equation}
We are used to these two quantities being shown as proportional to one another. This is not the case in general. It only applies when we restrict $\bm{k}$-space to a narrow-band of amplitudes.

\vspace{5pt}{\bf Localization}
The problem of localization was extensively explored by Mandel~\cite{mandel1966configuration}. We have summarized his arguments in Appendix~\ref{AppMandel}. The main result is that it is possible to define \emph{local} energy and photon number quantities by restricting the integrations of \cref{H,N} to a small volume $V$ instead of the full space. However, these local quantities make sense only when all dimensions of the volume are larger than the smallest wavelength being detected.

We can formalize this restriction by
\begin{itemize}
\item excluding the ball of radius $\kappa$ centered around $\bm{k}=\bm{0}$ from the domain of integration in $\bm{k}$-space in \cref{Sipe,Mandel};
\item dividing real space in a series of \emph{non-overlapping} small volumes $V_n$ with all dimensions larger than $2\pi/\kappa$;
\item limiting the integrations in \cref{H,N} to each small volume $V_n$ individually. This yields the local operators $\widehat{H}_{\kappa,n}(t)$ and $\widehat{N}_{\kappa,n}(t)$.
\end{itemize}

We can also use the first step above to define restricted wavefunctions $\bm\Psi_{\kappa}$ and $\bm\Phi_{\kappa}$. The expected energy absorption rate of an ideal bolometer encompassing $V_n$ and sensitive to all wavelengths below $2\pi/\kappa$ is then
\begin{equation}
    \dot{E}_{\kappa,n}(t)=\int_{V_n}\dd^3\bm{r}\;\pdv{\norm{\braket{\bm{r},t}{\bm\Psi_{\kappa}}}^2}{t},\label{Hobs}
\end{equation}
while the expected click rate on an ideal wideband photodetector encompassing $V_n$ and sensitive to all wavelengths below $2\pi/\kappa$ is
\begin{equation}
    \dot{N}_{\kappa,n}(t)=\int_{V_n}\dd^3\bm{r}\;\pdv{\norm{\braket{\bm{r},t}{\bm\Phi_{\kappa}}}^2}{t}.\label{Nobs}
\end{equation}

\vspace{5pt}{\bf Conclusion.}
There exists a direct link between Maxwell's equations and a photon density wavefunction. The second quantized amplitude of this wavefunction is Mandel's detection operator. Mandel himself noted that photons are approximately localizable, although not infinitely. We have shown that this approximate localizability is applicable to energy and photon densities equally. It can be argued that energy density is most relevant in momentum/frequency phase space, while photon density predicts quasi-instantaneous clicks on a photodetector and is thus most relevant in real space/time, which is the framework of QED.

\subsection*{Acknowledgements}
\vspace{-10pt}
I thank Nicolas Godbout, Bertrand Reulet, Denis Seletskiy, Cl\'ement Virally and Paul Virally for fruitful discussions leading to the completion of this project.
This work is funded in part by the Mid-Infrared Quantum Technology for Sensing (MIRAQLS)
Project, through the European Union’s Horizon Europe research and innovation programme under Agreement Number 101070700.

\bibliographystyle{apsrev4-2}
\bibliography{MaxwellSchroedinger}

\appendix
\vspace{-10pt}
\section{Matrix representations of Lie algebras}\label{ObviousL}
\vspace{-10pt}
The Pauli vector $\bm{\sigma}=\qty(\sigma_x,\sigma_y,\sigma_z)$ is the hermitian form of the $2\times2$ matrix representation of $\mathfrak{su}(2)$, the Lie algebra associated with spin-1/2 particles~\cite{pauli1927,hall_lie_2015,sakuraiQM}.
The components of the Pauli vector are
\begin{equation}
    \begin{matrix}
    \sigma_x\!=\!\begin{bmatrix}0&1\\1&0\end{bmatrix},&
    \sigma_y\!=\!\begin{bmatrix}0&i\\-i&0\end{bmatrix},&
    \sigma_z\!=\!\begin{bmatrix}1&0\\0&-1\end{bmatrix}.
    \end{matrix}
\end{equation}

The vector $\bm{L}=\qty(L_x,L_y,L_z)$ is the hermitian form of the $3\times3$ matrix representation of $\mathfrak{so}(3)$, the the Lie algebra associated with spin-1 particles~\cite{hall_lie_2015,sakuraiQM}. Its components are
\begin{equation}
    \begin{matrix}
    L_x\!=\!\scalemath{0.96}{\begin{bmatrix}0&0&0\\0&0&-i\\0&i&0\end{bmatrix}},&
    L_y\!=\!\scalemath{0.96}{\begin{bmatrix}0&0&i\\0&0&0\\-i&0&0\end{bmatrix}},&
    L_z\!=\!\scalemath{0.96}{\begin{bmatrix}0&-i&0\\i&0&0\\0&0&0\end{bmatrix}}.
    \end{matrix}
\end{equation}

\section{Transverse vector potential}\label{ObviousPotential}
\vspace{-10pt}
We write the potential vector in the Coulomb gauge as a sum of a positive and a negative frequency parts, $\bm{\mathcal{A}}=\bm{\mathcal{A}}^{(+)}+\bm{\mathcal{A}}^{(-)}$, with 
\begin{equation}
    \begin{split}
    \braket{\bm{r},t}{\bm{\mathcal{A}}^{(\pm)}}=&\int_{\mathbb{R}^3}\frac{\dd^3\bm{k}}{(2\pi)^{3/2}}\;\sum_{\sigma}\;\sqrt{\frac{\hbar\,Z_0}{2\,\abs{\bm{k}}}}\\&
    A^{(\pm)}_\sigma(\bm{k})\,e^{i(\bm{k}\cdot\bm{r}\mp\abs{\bm{k}}ct)}\;\tilde{\bm{u}}_\sigma(\bm{k}).
    \end{split}\label{A}
\end{equation}
Here, $Z_0=\sqrt{\mu_0/\epsilon_0}$ is the impedance of the vacuum.

For each $\bm{k}$, the unit vectors $\textstyle{\qty{\tilde{\bm{k}},\tilde{\bm{u}}_+(\bm{k}),\tilde{\bm{u}}_-(\bm{k})}}$ form an orthonormal basis, with $\tilde{\bm{u}}_+$ and $\tilde{\bm{u}}_-$ vectors corresponding to right-circular and left-circular helicity, respectively (see below).

\section{Right and left-circular basis vectors}\label{ObviousHelicity}
\vspace{-10pt}
For each $\bm{k}\neq\bm0$ vector in $\bm{k}$-space, we define the unit vector $\tilde{\bm{k}}=\bm{k}/\abs{\bm{k}}$, an arbitrary unit vector $\tilde{\bm{u}}_1$ orthogonal to $\bm{k}$, and the unit vector $\tilde{\bm{u}}_2=\tilde{\bm{k}}\cross\tilde{\bm{u}}_1$. Finally, we define the ``right-circular'' vector
\begin{equation}
    \tilde{\bm{u}}_+=\frac{\tilde{\bm{u}}_1+i\,\tilde{\bm{u}}_2}{\sqrt{2}},
\end{equation}
and the ``left-circular'' vector $\tilde{\bm{u}}_-=\tilde{\bm{k}}\cross\tilde{\bm{u}}_+$.

\section{Electric displacement and magnetic field}\label{ObviousFields}
\vspace{-10pt}
In the Coulomb gauge, where the potential is transverse, the transverse electric displacement and magnetic field are
\begin{center}
\vspace{-10pt}
\begin{subequations}
    \begin{minipage}{0.21\textwidth}
        \begin{equation} \label{AtoE}
        \bm{\mathcal{D}}=-\epsilon_0\pdv{\bm{\mathcal{A}}}{t},
    \end{equation}
\end{minipage}%
\begin{minipage}{0.27\textwidth}
    \begin{equation} \label{AtoH}
        \bm{\mathcal{B}}=\curl\bm{\mathcal{A}}.
    \end{equation}
\end{minipage}%
\end{subequations}
\end{center}

\section{Commutation relations}\label{ObviousCommutation}
\vspace{-10pt}
The usual commutation relations for the ladder operators of \cref{Q2} are
\vspace{-5pt}
\begin{equation}
    \begin{matrix}
    \comm{\hat{a}_{\sigma,\bm{k}}}{\hat{a}_{\sigma',\bm{k}'}}=\comm{\hat{a}^\dagger_{\sigma,\bm{k}}}{\hat{a}^\dagger_{\sigma',\bm{k}'}}=0;\\
    \comm{\hat{a}_{\sigma,\bm{k}}}{\hat{a}^\dagger_{\sigma',\bm{k}'}}=\delta_{\sigma,\sigma'}\;\delta(\bm{k}-\bm{k}').
    \end{matrix}
\end{equation}

\section{Creation-like operators}\label{ObviousCreation}
\vspace{-10pt}
The creation-like hermitian conjugates to the annihilation-like operators of \cref{Sipe,Mandel} are
\begin{equation}
    \widehat{\bm\psi}^\dagger_\sigma(\bm{r},t)=\int_{\mathbb{R}^3}\frac{\dd^3\bm{k}}{(2\pi)^{3/2}}\,\sqrt{\hbar\,\abs{\bm{k}}c}\;
    \hat{a}^\dagger_{\sigma,\bm{k}}\,e^{i(\bm{k}\cdot\bm{r}+\abs{\bm{k}}ct)}\;\tilde{\bm{u}}_\ell(\bm{k});\label{SipeDagger}
\end{equation}
\begin{equation}
    \widehat{\bm\phi}^\dagger_\sigma(\bm{r},t)=\int_{\mathbb{R}^3}\frac{\dd^3\bm{k}}{(2\pi)^{3/2}}\;
    \hat{a}^\dagger_{\sigma,\bm{k}}\,e^{i(\bm{k}\cdot\bm{r}+\abs{\bm{k}}ct-\pi/4)}\;\tilde{\bm{u}}_\ell(\bm{k}).\label{MandelDagger}
\end{equation}

\section{Mandel localization argument}\label{AppMandel}
\vspace{-10pt}
In his paper~\cite{mandel1966configuration}, Mandel gives a solid argument for the non-localization of photons in volumes with smallest linear dimension larger than the wavelengths being measured. We provide his reasoning here as a reference.

Mandel starts with the commutator
\begin{equation}
    \begin{split}
    &\comm{\hat{\bm\phi}_\ell(\bm{r},t)}{\hat{\bm\phi}^\dagger_{\ell'}(\bm{r}',t')}=\\
    &\delta_{\ell,\ell'}\int_{\mathbb{R}^3}\frac{\dd^3\bm{k}}{(2\pi)^3}\,e^{i[\bm{k}\cdot(\bm{r}-\bm{r}')-\abs{\bm{k}}c(t-t')]}\;\tilde{\bm{u}}_\ell(\bm{k}),
    \end{split}
\end{equation}
where we used the notations of this paper for clarity.

Calculating the commutator $\comm{\hat{\bm\phi}_\ell(\bm{r},t)}{\hat{N}_{V,\ell}(t')}$ over the cubic volume $V=L_x\times L_y\times Lz$ requires considering the integral
\begin{equation}
    \int_V\dd^3\bm{r}'\;e^{i(\bm{k}'-\bm{k})\cdot\bm{r}'}\propto\prod_{w\in\qty{x,y,z}}\mathrm{sinc}\qty[\frac{(k'_w-k_w)L_w}{2}].
\end{equation}
This quantity is essentially $\delta(\bm{k}'-\bm{k})$, or $\delta_{\bm{k},\bm{k}'}$ in the discrete case considered by Mandel, when the norms of the $\bm{k}$ vectors are large compared to the inverse of the smallest $L_w$. Mandel argues that $\hat{N}_V$ only acquires a meaningful sense in that limit. Hence, photons cannot be properly localized in volumes with all linear dimensions larger than the smallest wavelength being measured. This is in fact a direct consequence of Heisenberg's uncertainty principle~\cite{Heisenberg1927}.

The exact same problem arises when calculating the commutator $\comm{\hat{\bm\psi}_\ell(\bm{r},t)}{\hat{H}_{V,\ell}(t')}$. Hence, the localization of energy of the EM field suffers from the same limitations. 

\section{Energy $\leftrightarrow$ photon transforms}\label{Transforms}
\vspace{-10pt}
The causal transform defined in \cref{T} allows energy operators to be turned into photon operators. To be complete, there is also an anti-causal form
\begin{equation}
    \mathcal{T}_-[\bm{\mathcal{F}}](t)=\sqrt{\frac{2}{\hbar}}\int_{-\infty}^{0}\dd\tau\;\frac{\bm{\mathcal{F}}(t-\tau)}{\sqrt{-\tau}},\label{T-}
\end{equation}
and two interesting non-causal ``quadrature'' forms $\mathcal{T}_x=\qty(\mathcal{T}_++\mathcal{T}_-)/\sqrt{2}$ and $\mathcal{T}_y=\qty(\mathcal{T}_+-\mathcal{T}_-)/\sqrt{2}$, or
\begin{equation}
    \begin{matrix}
    \mathcal{T}_x[\bm{\mathcal{F}}](t)=\sqrt{\frac{1}{\hbar}}\int_{-\infty}^{+\infty}\dd\tau\;\frac{\bm{\mathcal{F}}(t-\tau)}{\sqrt{\abs{\tau}}},\\
    \mathcal{T}_y[\bm{\mathcal{F}}](t)=\sqrt{\frac{1}{\hbar}}\int_{-\infty}^{+\infty}\dd\tau\;\mathrm{sign}(\tau)\;\frac{\bm{\mathcal{F}}(t-\tau)}{\sqrt{\abs{\tau}}}.
    \end{matrix}
\end{equation}
All these transforms remove the $\sqrt{\hbar\,\abs{k}c}$ factor from the integrals. The causal and anti-causal transforms $\mathcal{T}_\pm$ add a $\pm\pi/4$ phase to the $\hat{a}_{\ell,\bm{k}}$ (and a $\mp\pi/4$ phase to their hermitian conjugates). The transform $\mathcal{T}_x$ adds no phase, while the $\mathcal{T}_y$ transform adds a $\pi/2$ phase, which is why we refer to them as ``quadrature'' transforms.

There also exist a set of transforms to turn photon operators to energy operators. They are
\vspace{-5pt}
\begin{equation}
    \begin{matrix}
    \overline{\mathcal{T}}_+[\bm{\mathcal{F}}](t)=-\sqrt{\frac{\hbar}{2}}\int_0^{+\infty}\dd\tau\;\frac{{\bm{\mathcal{F}}}(t-\tau)}{\tau^{3/2}},\\
    \overline{\mathcal{T}}_-[\bm{\mathcal{F}}](t)=-\sqrt{\frac{\hbar}{2}}\int_{-\infty}^{0}\dd\tau\;\frac{{\bm{\mathcal{F}}(t-\tau)}}{\qty(-\tau)^{3/2}},
    \end{matrix}
\end{equation}
in addition to the inverse quadrature transforms $\overline{\mathcal{T}}_x=\qty(\overline{\mathcal{T}}_++\overline{\mathcal{T}}_-)/\sqrt{2}$ and $\overline{\mathcal{T}}_y=\qty(\overline{\mathcal{T}}_+-\overline{\mathcal{T}}_-)/\sqrt{2}$, or
\begin{equation}
    \begin{matrix}
    \mathcal{T}_x[\bm{\mathcal{F}}](t)=-\sqrt{\frac{1}{\hbar}}\int_{-\infty}^{+\infty}\dd\tau\;\frac{\bm{\mathcal{F}}(t-\tau)}{\abs{\tau}^{3/2}},\\
    \mathcal{T}_y[\bm{\mathcal{F}}](t)=-\sqrt{\frac{1}{\hbar}}\int_{-\infty}^{+\infty}\dd\tau\;\mathrm{sign}(\tau)\;\frac{\bm{\mathcal{F}}(t-\tau)}{\abs{\tau}^{3/2}}.
    \end{matrix}
\end{equation}

\end{document}